\documentstyle[preprint,aps]{revtex}
\begin{document}

\tighten

\draft

\title{Comment on ``Critical behavior
of the chain--generating\\
 function of self--avoiding walks on\\
 the 
Sierpinski gasket family: The Euclidean limit''}
\author{Sava Milo\v sevi\'c}
\address{Faculty of Physics, University of
Belgrade, P.O.Box 368,\\ 11001 Belgrade, Serbia, Yugoslavia}
\author{Ivan \v Zivi\'c}
\address{Faculty of Natural Sciences and
Mathematics, University of Kragujevac,\\ 34000 Kragujevac, Serbia, Yugoslavia}
\author{Sun\v cica Elezovi\'c--Had\v zi\'c}
\address{Faculty of Physics, University of
Belgrade, P.O.Box 368,\\ 11001 Belgrade, Serbia, Yugoslavia}
\date{\today}

\maketitle
\begin{abstract} 
We refute  the claims made by Riera and
Chalub [Phys.~Rev.~E {\bf 58}, 4001 (1998)] by
demonstrating that they have not provided enough data
(requisite in their series expansion method) to draw reliable
conclusions about criticality of self-avoiding walks on the
Sierpinski gasket family of fractals.
\end{abstract}


\section*{}
The self--avoiding walk (SAW) on a lattice is a random walk
that must not contain self--intersections. The criticality
of SAWs has been extensively studied as a challenging
problem in statistical physics on the Euclidean lattices
and on fractal lattices as well. Accordingly, the question
has been posed whether the critical behavior of SAWs on a
Euclidean lattice can be retrieved via a limit of infinite
number of fractals whose properties gradually acquire the
corresponding Euclidean values. In this Comment we
scrutinize the methods used so far to answer the foregoing
question. 

The most frequently studied infinite family of finitely
ramified fractals appears to be the Sierpinski gasket (SG)
family. Each member of the SG family is labeled by an
integer $b$ ($2\le b\le\infty$), and when $b\to\infty$ both
the fractal $d_f$ and spectral $d_s$ dimension approach the
Euclidean value 2. Concerning the study of the criticality of
SAWs, these fractal lattices are perfect objects for application
of the renormalization group (RG) method, due to their
intrinsic dilation symmetry (the so-called self--similarity)
and their finite ramification. The latter property enables
one to construct a finite set of the RG transformations and
therefrom an exact treatment of the problem. This treatment
was first applied by Dhar \cite{Dhar}, for $b=2$, and later it was
extended \cite{EKM} up to $b=8$. The obtained results of the
corresponding critical exponents, for the finite sequence
$2\le b\le8$, were not sufficient to infer their relation to
the relevant Euclidean values. However, these results
inspired the finite--size scaling (FSS) approach to the
problem \cite{FSS}, which brought about the prediction that the
SAW critical exponents on fractals do not necessarily
approach their Euclidean values when $b\to\infty$. Indeed,
Dhar \cite{FSS} found that the critical exponent $\nu$, associated
with the SAW end--to--end distance, tends to the Euclidean
value 3/4, whereas the critical exponent $\gamma$,
associated with the total number of distinct SAWs,
approaches 133/32, being always larger than the Euclidean
value $\gamma_E=43/32$.

The intriguing FSS results motivated endeavors to extend
the exact RG results beyond $b=8$. However, since this
extension appeared to be an arduous task, a new insight was
needed. This insight came from a formulation of the Monte
Carlo renormalization group (MCRG) method for fractals
[4,5], which produced values for $\nu$ and $\gamma$ up to
$b=80$. For $2\le b\le8$ the MCRG findings deviated, from
the exact values, at most 0.03\%, in the case of $\nu$, and
0.2\% in the case of $\gamma$. In addition, the behavior of
the entire sequence of the MCRG findings, as a function of
$b$, supported the FSS predictions.

Recently, Riera and Chalub [6] made a different type of
endeavor to obtain results for the critical exponent
$\gamma$ for large $b$, by applying an original series
expansion method [7]. However, the data of Riera and Chalub
(RC) display a quite different behavior than the MCRG
results (see Fig. 1). As regards comparison of the RC
results with the available exact RG results [2], one may
notice a surprising discrepancy: for $b=7$ the RC result deviates
19\% (which should be compared with the respective MCRG
deviation 0.13\%), while for $b=8$ the RC result deviates
33\% from the exact result (which is again much larger than
the corresponding MCRG deviation 0.15\%). On the other
hand, concerning behavior of $\gamma$ beyond $b=8$, the RC
results start to decrease, whereas the MCRG results
monotonically increase.  Furthermore, Riera and Chalub [6]
claimed that, in contrast with the FSS prediction [3],
$\gamma$ should approach the Euclidean value 43/32=1.34375,
in the limit of very large $b$. These discrepancies call
for inspection of both methods, that is, of the MCRG
technique [4,5] and the series expansion method [6,7]. We
are going first to reexamine our MCRG approach, and then we
shall comment on the applicability of the  RC series
expansion method for large $b$.

We have found that the best way to check the validity of
the MCRG method, for large $b$, is to apply it in a case of
a random walk model that is exactly solvable for all
possible $b$. To this end, the so-called piecewise directed
walk (PDW) [8,9] turned out to be quite appropriate. The
PDW model describes such a random self-avoiding walk on SG
fractals in which the walker is allowed to choose randomly,
but self-similarly, limited number of possible step
directions [8]. This model corresponds to the directed
random walk on Euclidean lattices, in which case $\nu_E=1$
and $\gamma_E=1$. By applying the exact RG approach the
critical exponent $\nu$ and $\gamma$ for PDW have been
obtained [8,9] for each $b$ ($2\le b<\infty$). Moreover, it was
demonstrated exactly that $\nu$ approaches the Euclidean
value $\nu_E=1$, while $\gamma$ tends to non-Euclidean value
$\gamma=2$, when $b\to\infty$. Here, we apply the MCRG
method (used in the case of SAW in [4,5]) to calculate 
$\nu$ and $\gamma$ of the PDW model for $2\le b\le100$. Our
results, together with the exact findings, are presented in
Table~I and depicted in Fig.~2 and Fig.~3. One can see that,
in the entire region under study, the agreement between the
MCRG results and the exact data is excellent. Indeed, the
deviation of the MCRG results for $\nu$ from the
corresponding exact results is at most 0.08\%, while in the
case of $\gamma$ it is at most 0.8\%. This test of the MCRG
method provides novel reliability for its application
in studies of random walks on SG with large $b$. 

Because of the confirmed reliability of the MCRG method,
and because Riera and Chalub [6] have obtained quite
different results, in the case of SAWs on SG, we have
reason to assume that their conclusions were obtained in a
wrong way. Thus, we may pose a question what was wrong in
the application of the series expansion method in the work
of Riera and Chalub [6]. Let us start with mentioning that
in the series expansion study of SAW the first task is to
determine the number $c_n(b)$ of all possible SAWs for a
given number $n$ of steps, where $1\le n\le n_{max}$. Of
course, in practice, it is desirable to perform this
enumeration for very large $n_{max}$, as the corresponding
numbers $c_n(b)$ of all $n$-step SAWs represent
coefficients of the relevant generating function
$C_b(x)=\sum_{n=1}^{\infty}c_n(b)x^n$ (where $x$ is the
weight factor for each step), whose singular behavior
determines critical exponents of SAWs. In order to take
into account existence of the SG lacunarity, the average
end-to-end distance of the set of $n$-step SAWs should be
larger than the size of the smallest homogeneous part of
the SG fractal [3], that is, $n_{max}$ should be larger
than $b^{4/3}$. In a case when the number of steps is
smaller than $b^{4/3}$ the corresponding SAWs percieve the
underlying fractal lattice as a Euclidean substrate.  In
order to make it more transpicuous, we present in Fig.~4
the curve $n=b^{4/3}$ which divides the ($b,n$) plane in
two regions so that one of them corresponds to the fractal
behavior of $n$-step SAWs, while the other corresponds to
the Euclidean behavior. In the same figure, for a given
$b$, we also depict the number of coefficients (empty small
triangles) that were obtained by Riera and Chalub [6] for
the corresponding SG fractal, in their series expansion
approach. One can see that only for $2\le b\le 8$  Riera
and Chalub generated sufficient number of coefficients
$c_n(b)$ so as to probe the fractal--behavior region (in
which the condition $n_{max}\ge b^{4/3}$ is satisfied). On
the other hand, for $b>8$, in all cases studied, the
maximum length $n_{max}$ of enumerated SAWs [6] is not
larger than 16, and thereby the corresponding generating
functions $C_b(x)$ remain in the domain of the Euclidean
behavior (see Fig.~4).  For instance, in order to study
criticality of SAWs on the SG fractal, for $b=80$, it is
prerequisite to calculate all coefficients $c_n(b)$ in the
interval $1\le n\le n_{max}$, where $n_{max}$ must be
larger than $80^{4/3}\approx339$, which is far beyond
$n_{max}=13$ that was reached in [6] for $b=80$. This
explains why the RC results for the SG critical exponent
$\gamma$ (see Fig.~1), with increasing $b$, wrongly become
closer to the Euclidean value 1.34375.

The problems discussed above, that is, the problems with
not long enough SAWs, do not appear in the MCRG study of
SAWs on the SG fractals, as the RG method in general takes
into account SAWs of all length scales. Incidentally, we
would like to mention that in [6] it was erroneously quoted
that in the MCRG studies [4,5] one Monte Carlo (MC)
realization corresponds to simulation of one SAW. In fact,
one MC realization implies simulations of all possible
walks on the fractal generator, which appears to be the
smallest homogeneous part of the SG fractal. For instance,
for the $b=80$ SG fractal, in order to calculate the
critical exponent $\gamma$, in one MC realization we
simulated [5] all $n$-step SAWs with $n$ ranging between 1
and 3240.

Finally, we would like to comment on the analytical
argument, given in [6], which was assumed to support the
claim $\lim_{b\to\infty}\gamma=\gamma_E$.  One can observe
that the corresponding argument does not exploit particular
properties of the SAWs studied.  Thus, if the argument were
valid, it could be applied to other types of SAWs on
fractals leading to the same conclusion
$\lim_{b\to\infty}\gamma=\gamma_E$.  However, the case of
the PDW discussed in this comment is a definite
counterexample to the foregoing conclusion, as it was
rigorously demonstrated [8,9] that in this case
$\lim_{b\to\infty}\gamma\not=\gamma_E$.

In the conclusion, let us state that in this comment we
have vindicated that the MCRG technique for studying
the SAW critical exponents on fractals is a reliable method
and a valuable tool in discussing the query whether the
critical behavior of SAWs on a Euclidean lattice can be
achieved through a limit of infinite number of fractals
whose properties gradually acquire the corresponding
Euclidean values. On the other hand, we have demonstrated
that Riera and Chalub [6], in an attempt to answer the
mentioned query by applying the series expansion method,
have not provided sufficient number of numerical data for a
study of criticality of SAWs on the SG fractals with finite
scaling parameters $b$. Therefore, any inference from such
a set of data about the large $b$ behaviour of the critical
exponent $\gamma$ cannot be tenable.

We would like to acknowledge helpful and inspiring
correspondence with D.~Dhar concerning the matter discussed
in this Comment.

\vfill
\eject

\begin{figure}
\caption{The SAW critical exponent $\gamma$ as a function of
$1/b$. The solid triangles represent results of the MCRG
calculation [5], while the open triangles represent results
obtained by Riera and Chalub (RC) via the series expansion
method [6]. In both cases the solid lines that connect the
data symbols serve as the guide to the eye. One should
observe unusually large error bars in the case of RC
results, whereas in the case of the MCRG results the error
bars lie within the data symbols. The horizontal dashed
line represents the Euclidean value $\gamma_E=43/32$ (which
is also indicated by the solid horizontal arrow).}
\label{fig1}
\end{figure}

\begin{figure}
\caption{ The PDW critical exponent $\nu$ as a function  of
$1/b$. The solid line represents exact results [8], while
the solid triangles represent the MCRG results (see
Table~I). The error bars for the MCRG results lie within
the data symbols. The inset is given in order to depict the
limiting value of $\nu$ (solid circle) which coincides with
Euclidean value $\nu_E=1$.}
\label{ni}
\end{figure}

\begin{figure}
\caption{ The PDW critical exponent $\gamma$ as a function of
$1/b$. The solid line represents exact results [9], while
the solid triangles represent the MCRG results (see
Table~I). The data error bars for the MCRG results for $b$
up to 30 are invisible, while for larger $b$ the error bars
are comparable with the size of the symbols. The inset is
given in order to depict the limiting value  $\gamma=2$
(solid circle) which is different from the Euclidean value
$\gamma_E=1$.}
\label{gama}
\end{figure}

\begin{figure}
\caption{ The schematic representation of the fractal and
Euclidean behavior of SAWs on the SG fractals. Regions of
the two different behaviors are separated by the solid line
$n=b^{4/3}$. The height of each vertical line (comprised of
small triangles) corresponds to the maximum length of the
$n$-step SAWs enumerated in the series expansion approach
by Riera and Chalub [6], for a given $b$. In order to study
criticality of SAWs on the SG fractals it is necessary to
extend the vertical lines beyond the solid curve. Accordingly,
one should observe that results of Riera and Chalub [6] do
not probe the fractal region for $b>8$.}
\label{scheme}
\end{figure}

\begin{table}
\caption{The exact RG and MCRG (in the brackets) results of
the PDW critical exponents $\nu$ and $\gamma$ for the SG
fractals for $2\le b\le100$. The exact RG results have been
obtained in [8,9], while the MCRG results are calculated in
the present work.}
\begin{tabular}{ccc}
$b$&{$\nu$} exact(MC)&{$\gamma$} exact(MC)\\
\tableline
2  & 0.79870 (0.79801$\pm$0.00075)  & 0.9520 (0.9503$\pm$ 0.0018) \\
3  & 0.82625 (0.82683$\pm$0.00043)  & 0.9631 (0.9638$\pm$ 0.0014)  \\  
4  & 0.84311 (0.84332$\pm$0.00010) & 0.9673 (0.9665$\pm$ 0.0011)  \\
5  & 0.85469 (0.85470$\pm$0.00008)  & 0.9695 (0.9690$\pm$ 0.0011)  \\
6 & 0.86329 (0.86329$\pm$0.00006)  &0.9711 (0.9707$\pm$0.0012)   \\
7 & 0.87000  (0.86992$\pm$0.00006)  &0.9726  (0.9723$\pm$0.0012)   \\
8& 0.87542  (0.87536$\pm$0.00005)  &0.9742  (0.9732$\pm$0.0012)   \\
9 & 0.87992  (0.87978$\pm$0.00004)  &0.9759  (0.9749$\pm$0.0013)   \\
10& 0.88374  (0.88370$\pm$0.00013)  &0.9777  (0.9770$\pm$0.0014)   \\
12& 0.88992  (0.88984$\pm$0.00011) &0.9815\ (0.9812$\pm$0.0015)   \\
15& 0.89679  (0.89662$\pm$0.00009) &0.9877\ (0.9839$\pm$0.0016)   \\
17& 0.90035  (0.90040$\pm$0.00008) &0.9919  (0.9911$\pm$0.0016)  \\
20& 0.90467  (0.90460$\pm$0.00007)  &0.9982  (0.9946$\pm$0.0018)   \\
25& 0.91011  (0.91002$\pm$0.00006)  &1.0084  (1.0005$\pm$0.0019)   \\
30& 0.91417  (0.91415$\pm$0.00002)  &1.0180  (1.0181$\pm$0.0020)   \\
35& 0.91736  (0.91727$\pm$0.00005)  &1.0270  (1.0194$\pm$0.0022)   \\
40& 0.91996  (0.91995$\pm$0.00005)  &1.0354  (1.0368$\pm$0.0023)   \\
50  & 0.92399   (0.92393$\pm$0.00004) & 1.0504  (1.0457$\pm$ 0.0025)  \\
60 & 0.92701  (0.92704$\pm$0.00003) &1.0634  (1.0696$\pm$0.0026)  \\
70 & 0.92940  (0.92934$\pm$0.00003)  & 1.0750  (1.0789$\pm$ 0.0028)  \\      
80& 0.93136  (0.93136$\pm$0.00001)  &1.0853  (1.0854$\pm$0.0029)   \\
90& 0.93300  (0.93300$\pm$0.00001)  &1.0945  (1.0948$\pm$0.0031)   \\
100&0.93441  (0.93437$\pm$0.00002)  &1.1029  (1.1089$\pm$0.0032)   \\
\end{tabular}
\end{table}

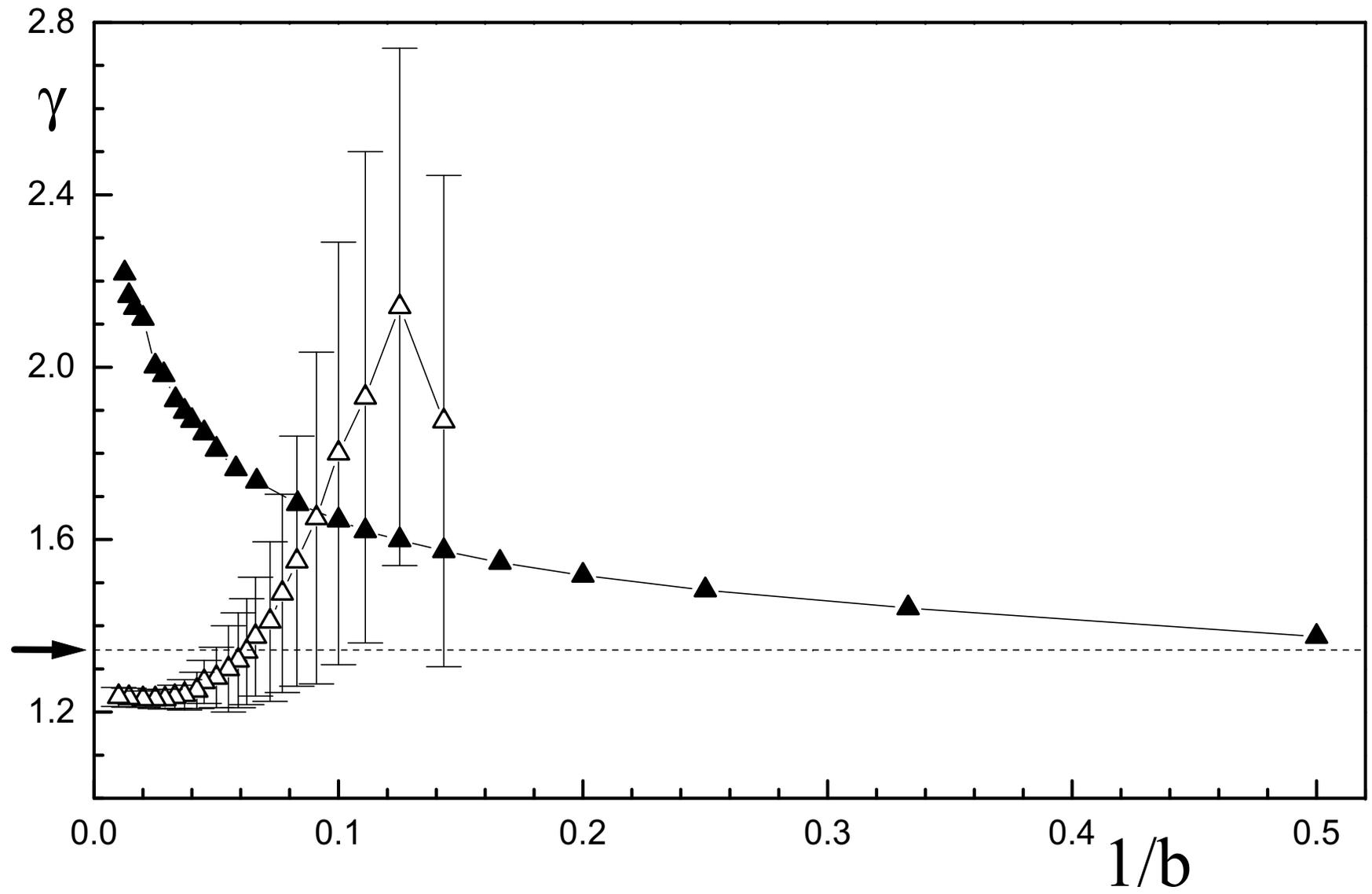
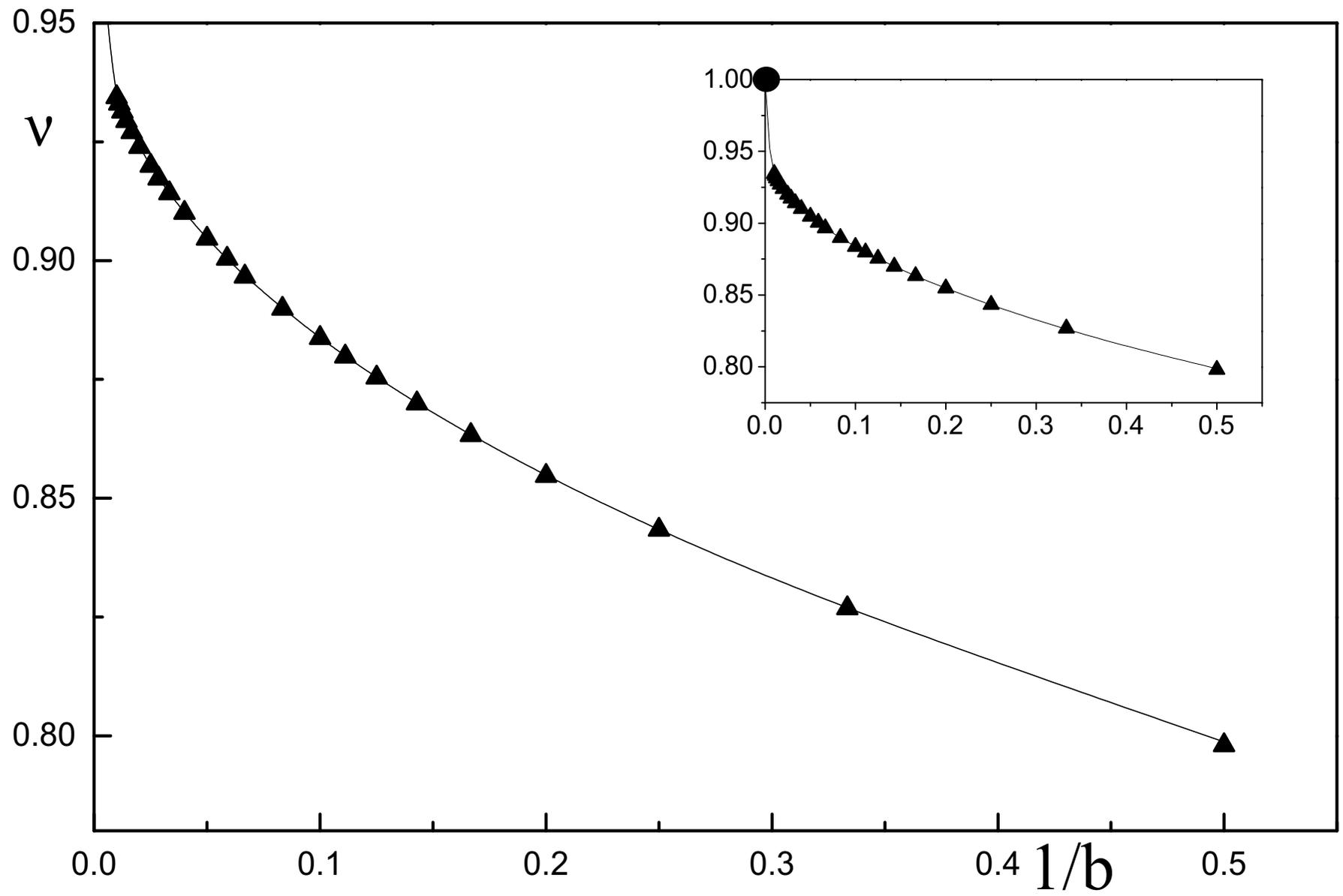
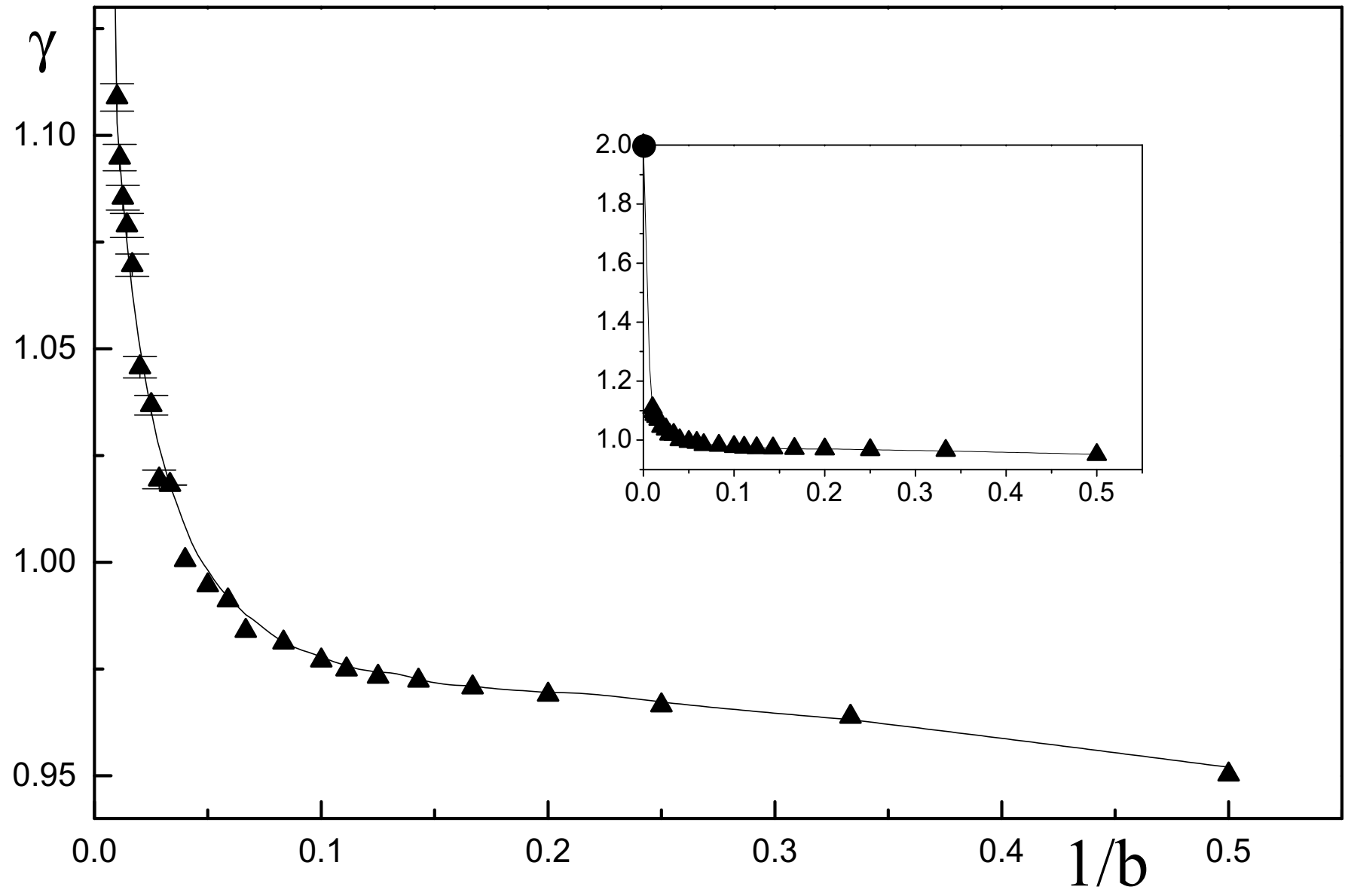
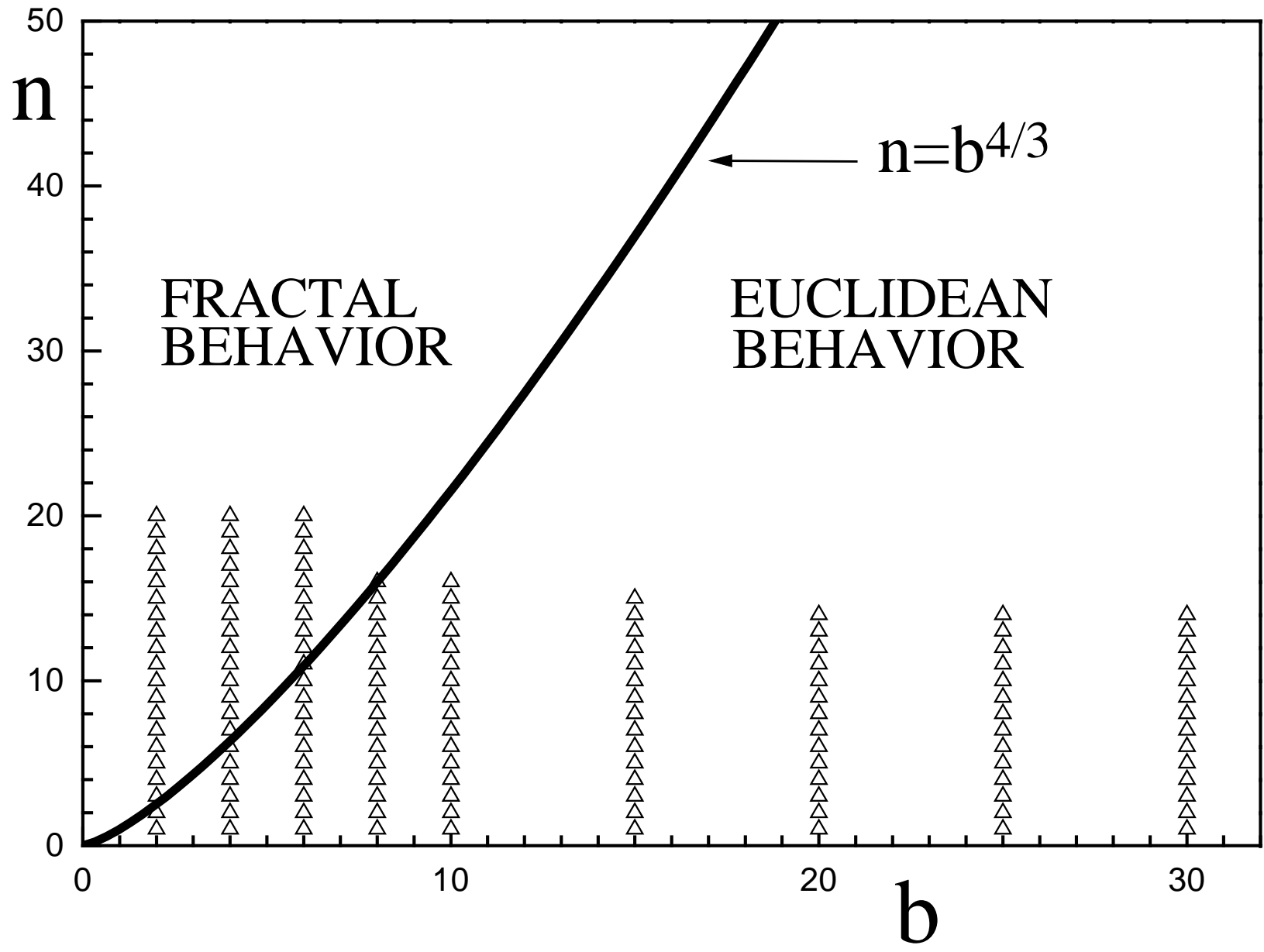
\end{document}